
\magnification=\magstep 1
\input amstex
\documentstyle{amsppt}
\NoRunningHeads
\tolerance=10000
\pagewidth{126mm}
\pageheight{200mm}
\hcorrection{2mm}
\overfullrule=0pt

\def\id{\operatorname{id}}

\def\Diff{\operatorname{Diff}}

\def\({\left(}
\def\){\right)}
\def\[{\left[}
\def\]{\right]}
\def\lbr{\left\{}
\def\rbr{\right\}}

\def\defn{\overset\text{def}\to=}
\def\wel2{\ell_{2,\text{w}}}
\def\ae{\overset\text{a.e.}\to=}
\def\ess{\operatornamewithlimits{ess{\phantom s} sup}}
\topmatter
\author {\rm Tze-Chuen Toh}
\footnote"$^*$"{e-mail:
tct105\@rsphy2.anu.edu.au\hfill\hfill} {\rm and Malcolm R. Anderson}
\footnote"$^\dag$"{e-mail:
m.anderson\@cowan.edu.au\hfill\hfill}
\endauthor
\title Knots And Classical 3-Geometries\endtitle
\abstract
It has been conjectured by Rovelli that there is {\sl a}
correspondence between the space of link classes of a Riemannian
3-manifold and the
space of 3-geometries (on the same manifold). An exact statement of his
conjecture will be established and then verified
for the case when the 3-manifold is compact, orientable and closed.
\medpagebreak

PACS. 04.60, 02.40.

Mathematics Subject Classications (1991). 53C80, 54P47 + 35.

(This article will appear in the January, 1995 edition of the Journal of
Mathematical Physics.)
\endabstract
\affil{\it ${}^*$Department of Theoretical Physics\\
Research School of Physical Sciences and Engineering\\
The Australian National University\\
Canberra ACT 0200 AUSTRALIA\\
and\\
${}^\dag$Department of Mathematics\\
Edith Cowan University\\
Mt. Lawley WA 6050 AUSTRALIA}\endaffil
\endtopmatter
\document
\baselineskip=16pt

\subheading{1. Introduction}
\medpagebreak

In \cite{6, p. 1661}, Rovelli sketched a proof showing how a certain
collection of $n$-loops, which he called {\it weaves}, are related to
the flat 3-metric. He then conjectured that perhaps
there exists a relationship between $n$-loops,
\footnote{ An {\it $n$-loop}
$\gamma\defn\{\, \gamma^1,\dots, \gamma^n\,\}$
is just a subset of the loop space consisting of $n$ loops; {\sl i.e.},
$\gamma^i$, for each $i=1,\dots, n$, are (distinct) closed curves
in $\Sigma$.}
for $n<\infty$, and 3-metrics. The relationship
between $n$-loops, for $n<\infty$, and 3-metrics will not be answered in
this paper (and so, it still remains an open question); however, what
will be shown in this article is that there exists a precise
relationship between 3-geometries and a subset of $\aleph_0$-knots,
where an {\it $n$-knot} is defined to be an equivalence class of $n$-loops
under (smooth) ambient isotopies ({\sl cf.} \S 3).
The approach given here is entirely
different to that outlined by Rovelli in \cite{6}: tersely, Rovelli
introduced a lattice spacing on the 3-manifold---the distance between
parallel non-intersecting curves which defines a weave in 3-space---to
obtain his conclusion regarding weaves and flat metrics; this,
in turn, motivated his conjecture. Here, no such assumptions will be
made and the results are purely `topological'.

The attention here will be focused on compact, Riemannian 3-manifolds.
In this paper, the term {\it Riemannian metric} means a smooth,
non-degenerate, symmetric, covariant 2-tensor that is {\sl
positive-definite} on the base manifold.
The fact that the 3-manifold is separable is crucial in the
construction: this, at least, explains why $\aleph_0$-loops are used
rather than $n$-loops for $n<\infty$.
The main interest in Rovelli's conjecture
is that it will provide a tentative physical
interpretation of the loop representation of quantum gravity \cite{7}:
it yields a possible insight into
the interweaving of topology and geometry at the quantum level. More will
be said in section 5.

In all that follows, the spatial (Riemannian)
3-manifold, denoted by $\Sigma$, is assumed
to be smooth, orientable,
closed and compact; $\Bbb R_+\defn \{\, s\in\Bbb R\mid s\geqq 0\,\}$ and
$I\defn [0,1]$. Lastly, let $\Diff^+(\Sigma)$ denote the group of
smooth, orientation-preserving, diffeomorphisms on $\Sigma$.
An overview of this paper runs as follows: section 2 introduces
the required notations and definitions, whilst in section 3, the
property of the space of $\aleph_0$-knots of a subset of
$\aleph_0$-loops will be examined. This space will establish the sought
for correspondence between topology and geometry. In
section 4, a variant of
the Rovelli conjecture will be formulated precisely and then
verified; then, in
section 5, some interesting speculations regarding
the results established in \S 4 will be
outlined.
\bigpagebreak

\subheading{2. Preliminary Definitions and Notations}
\medpagebreak

Let $\Omega_\Sigma=\{\, \gamma: I\to\Sigma\mid \gamma(0)=\gamma(1),\,
\gamma\text{ continuous }\}$ be the space of loops in $\Sigma$ and equip
it with the compact-open topology. Fix a Riemannian metric  $\hat q$ on
$\Sigma$ and let $\hat d:\Sigma\times\Sigma\to \Bbb R_+$ be the distance
function on $\Sigma$ induced by $\hat q$. Then, the metric $d_{\Omega}:
\Omega_\Sigma\times\Omega_\Sigma\to\Bbb R_+$ defined by
$$d_{\Omega}(\gamma,\eta)\defn \sup_{t\in I}\hat d(\gamma(t),\eta(t))$$
induces a metric topology on $\Omega_\Sigma$ which is compatible with
its compact-open topology \cite{1, p. 263, theorem 4.2.17}.
\medpagebreak

\remark{2.1. Remark} Since for each (admissible)
Riemannian metric $q$ on $\Sigma$,
\footnote{The topology on $\Sigma$
and its differentiable structure are
of course assumed fixed throughout the discussion.}
the $d_q$-topology---where $d_q$ is the distance function induced on
$\Sigma$ by $q$---coincides with the manifold topology, it follows that
all the metrics $d_q$ are equivalent to one another. Hence, all the
metrics $d_{\Omega}$ are also equivalent.
\endremark
\medpagebreak

Let $\tilde\Cal L_\Sigma\subset\Omega_\Sigma$ be the space of piecewise
smooth loops in $\Sigma$ endowed with the subspace topology. Next,
quotient away the constant loops---{\sl i.e.}, $\gamma(I)=\{x_\gamma\}$,
some $x_\gamma\in\Sigma$---in $\tilde\Cal L_\Sigma$ as follows. Define
an equivalence relation $\tilde\Cal R\subset\tilde\Cal L_\Sigma\times
\tilde\Cal L_\Sigma$ such that $\forall\, (\gamma,\eta)\in\tilde \Cal
R$, $\gamma$ and $\eta$ are constant loops in $\tilde\Cal L_\Sigma$. Let
$\Cal L_\Sigma\defn \tilde\Cal L_\Sigma/\tilde\Cal R$ be the quotient
space and $\tilde\pi:\tilde\Cal L_\Sigma\to \Cal L_\Sigma$ the
natural map. It is clear that $\tilde\Cal R$ is closed in $\tilde\Cal
L_\Sigma\times\tilde\Cal L_\Sigma$. Furthermore, observe from the
construction that if $\Cal L_0=\{\, \gamma\in\tilde\Cal L_\Sigma\mid
\gamma\text{ is a constant loop }\}$, then $\tilde\pi|( \tilde\Cal
L_\Sigma-\Cal L_0)$ coincides with the inclusion map $\tilde\Cal
L_\Sigma-\Cal L_0\hookrightarrow\tilde\Cal L_\Sigma$; that is,
$\tilde\pi|(\tilde\Cal L_\Sigma-\Cal L_0)= \id_{\tilde\Cal L_\Sigma}
|(\tilde\Cal L_\Sigma-\Cal L_0)$.

\proclaim{2.2. Lemma} $\Cal L_0$ is closed and nowhere dense in
$\tilde\Cal L_\Sigma$.
\endproclaim
\demo{Proof} Let $\{\gamma_n\}_n$ be a sequence in $\Cal L_0$ which
converges to $\gamma_0\in\tilde\Cal L_\Sigma$. By definition, $\gamma_n(I)=
x_n\in\Sigma\,\;\forall\,n$ and $\Sigma$ compact Hausdorff imply that
$\exists\,x_0\in\Sigma$ and a subsequence $\{x_{n_k}\}_k\subset
\{x_n\}_n$ such that $x_{n_k}\to x_0$. Since $\forall\,\varepsilon>0,
\exists\, N_\varepsilon>0$ such that $d_\Omega(\gamma_n,\gamma_0)=
\sup_{t\in I}\hat d(\gamma_n(t),\gamma_0(t))=\sup_{t\in I}\hat d(x_n,
\gamma_0(t))
< \varepsilon\,\;\forall\,
n>N_\varepsilon$, it follows
at once that $\gamma_0(t)\equiv x_0$ on $I$ and $\Cal L_0$
is thus closed, where $d_\Omega|\tilde \Cal L_\Sigma
\times\tilde\Cal L_\Sigma$ is denoted by
$d_\Omega$ for simplicity.

Finally, to complete the proof, suppose that the interior $\Cal
L_0^\circ\neq\varnothing$. Then, for any fixed $\eta\in\Cal L_0^\circ$, there
exists a neighbourhood $N_\eta= \bigcap_{i=1}^n M(K_i,O_i)$, for some
$n<\infty$---where $M(K_i,O_i)\defn \{\, \gamma\in\tilde\Cal L_\Sigma
\mid K_i\subset I\text{ is compact, }\gamma(K_i)\subset O_i,\,
O_i\subset\Sigma\text{ is
open }\}$---such that
$N_\eta\subset \Cal L_0^\circ$. However, $\eta\in
N_\eta$ and $\eta(I)=x_\eta$ for some $x_\eta\in \Sigma\Rightarrow
O\equiv\bigcap_{i=1}^nO_i\neq\varnothing$. Let $L_\eta=\{\,
\gamma\in\tilde\Cal L_\Sigma\mid
\gamma(I)\subset O\,\}-\Cal L_0$. Evidently,
$L_\eta\neq\varnothing$, and in particular,
$L_\eta\cap N_\eta\neq\varnothing$,
which is a contradiction. Hence, $\Cal L_0^\circ\equiv\varnothing$, as
required. $\qed$
\enddemo

The neighbourhood base of $0_\Sigma\equiv\tilde\pi(\gamma)\;\,\forall\,
\gamma\in\Cal L_0$ can now be constructed. It follows from the quotient
topology and lemma 2.2 that for each neighbourhood $N_{0_\Sigma}$ of
$0_\Sigma$, $\tilde\pi^{-1}(N_{0_\Sigma})$ must be a neighbourhood of $\Cal
L_0$. Hence, from the definition of $\tilde\pi$, the neighbourhood base
of $0_\Sigma$ consists precisely of subsets $\tilde\pi(N)$, where $N$ is
a neighbourhood of $\Cal L_0$ in $\tilde\Cal L_\Sigma$. Explicitly, a
neighbourhood of $0_\Sigma$ is of the form $\bigcup_{\eta\in\Cal L_0}
\tilde\pi(N_\eta)$, where $N_\eta$ is a neighbourhood of $\eta$ in
$\tilde\Cal L_\Sigma$.
Notice however,
that $\tilde\pi$ is not an open map. For given any neighbourhood
$N_\eta$ of $\eta\in\Cal L_0$, $\tilde\pi^{-1}\circ \tilde\pi(N_\eta)=
N_\eta\cup\Cal L_0$ which is neither closed nor open (by lemma 2.2) if $\Cal
L_0\not\subset N_\eta$. Nevertheless, for each neighbourhood $N$ of
$\Cal L_0$, $\tilde\pi(N)$ is a neighbourhood of $0_\Sigma$.

\proclaim{2.3. Lemma} $\tilde\pi:\tilde\Cal L_\Sigma\to\Cal L_\Sigma$ is
closed.
\endproclaim
\demo{Proof} To establish this claim, it is enough to show that
$\tilde\pi$ maps closed neighbourhoods $C_\eta$ of $\eta\in\Cal L_0$ into
closed neighbourhoods of $0_\Sigma$. Invoking the quotient topology, it
will suffice to verify that $\tilde\pi^{-1}\circ\tilde\pi (C_\eta)$ is
closed in $\tilde\Cal L_\Sigma$. Since $\tilde\pi^{-1}\circ\tilde\pi
(C_\eta)=C_\eta\cup\Cal L_0$ by definition, lemma 2.2 yields the
desired result. $\qed$
\enddemo

\proclaim{2.4. Theorem} $\Cal L_\Sigma$ is metrizable.
\endproclaim
\demo{Proof} By corollary 2.3, it will suffice to show that $\Cal
L_\Sigma$ is first countable \cite{1, p. 285, theorem 4.4.17}, and
from the definition of $\tilde\pi$,
it is enough to verify that $0_\Sigma$ has a countable
neighbhourhood base, since each $\gamma\in\Cal L_\Sigma- \{0_\Sigma\}$
has a countable neighbourhood base (by definition).
Let $\frak B_\eta =\{\,
B_{\frac 1n}(\eta)\mid n\in\Bbb N\,\}$
be a countable neighbourhood
base of $\eta\in\Cal L_0$ in $\tilde\Cal L_\Sigma$. Since, by
definition, each subset of the form $\bigcup_{\eta\in\Cal L_0}\tilde
\pi(B_{\frac 1n}(\eta))$ defines a neighbourhood of $0_\Sigma$, it
follows that the
collection $\frak B_{0_\Sigma}$ defined by
$$\frak B_{0_\Sigma}=\lbr \bigcup_{\eta\in\Cal L_0}\tilde
\pi(B_{\frac 1n}(\eta))
\,\Bigg|\Bigg.\, n\in
\Bbb N\rbr$$
is a countable neighbourhood base of $0_\Sigma$. Hence, $\Cal
L_\Sigma$ is metrizable, as claimed. $\qed$
\enddemo

In the following account,
call a curve in $\Sigma$ a
{\it $q$-geodesic} if it is a geodesic in
$\Sigma$ relative to the Riemannian metric $q$. Also, if $\gamma,\eta$
are curves such that $\gamma(1)=\eta(0)$, then define $\gamma*\eta$ by
$$\gamma*\eta(t)=\cases \gamma(2t)\qquad
\text{ for } 0\leqq t\leqq\frac
12,\\
\eta(2t-1)\,\,\text{ for }\frac 12\leqq t\leqq 1.\endcases
$$

\definition{2.5. Definition} Let $\gamma\in\Cal L_\Sigma$. Then,
$\gamma$ is said to be a {\it piecewise geodesic loop} if there exists a
Riemannian metric $q$ on $\Sigma$ and $n$ smooth
$q$-geodesics $\gamma_1,\dots,
\gamma_n: I\to\Sigma,\, 1\leqq n<\infty$,
such that $\gamma=\gamma_1*
\dots*\gamma_n$.\footnote{ Note that each
$\gamma_i$ in $\gamma$ is still
a $q$-geodesic with respect to its new parametrization $[\frac{i-1}{n},
\frac{i}{n}]$, as the
reparametrization $\gamma_i(t)\to \gamma_i(nt-i+1)\equiv \gamma|
[\frac{i-1}{n},\frac{i}{n}]$ is clearly an affine transformation.}
\enddefinition

Let $\Gamma^+_2$ denote the space of (smooth) Riemannian metrics on $\Sigma$
(endowed with the compact C${}^\infty$-topology
\footnote{The compact
C${}^\infty$-topology is defined in the appendix.})
and $D_\Sigma\subset\Sigma$ a countably dense subset of $\Sigma$.
Now, define $\Cal M_\infty[q]$,
for each $q\in\Gamma^+_2$, to be the set of
all countably infinite multi-loops $\gamma=\{\,\gamma^i\mid i\in\Bbb N\,\}$
satisfying the following two properties:
\roster
\item for each $i$, $\gamma^i\in\Cal L_\Sigma$
is a piecewise (affinely parametrized) $q$-geodesic loop in $\Sigma$,
\item the subset $\gamma$ is in bijective
correspondence with $D_\Sigma$ under the map $\gamma^i\mapsto
\gamma^i(0)$.
\endroster
Finally, set $\Cal
M_\infty[\Gamma^+_2]=\bigcup_{q\in\Gamma^+_2}\Cal M_\infty[q]$.
An immediate consequence of the definition is the following two
observations. Suppose $\gamma\in\Cal
M_\infty[q]\cap \Cal M_\infty[q']$. Let $\Gamma(q)$ and $\Gamma(q')$ be
the Riemannian connections of $q$ and $q'$ respectively. Fix an
admissible atlas $\{(U_\alpha,\psi_\alpha)\}_\alpha$ on $\Sigma$. Then,
with respect to each chart $U_\alpha$,
$$(\ddot\gamma_\alpha^i)^\ell+\Gamma_\alpha(q)^\ell_{kj}
(\dot\gamma_\alpha^i)^k
(\dot\gamma_\alpha^i)^j\ae 0\;\text{ and }\;
(\ddot\gamma_\alpha^i)^\ell+\Gamma_\alpha(q')^\ell_{kj}
(\dot\gamma_\alpha^i)^k
(\dot\gamma_\alpha^i)^j\ae 0$$
on $\gamma^i(I)\cap U_\alpha$ for each $i$ (no
summation over $\alpha$, obviously), where $F(t)\ae 0$ means $F(t)=0$ on
$I$ {\sl apart} from a {\it finite} number of points in $I$.
Hence, $(\Gamma_\alpha(q)^\ell_{kj}-\Gamma_\alpha(q')^\ell_{kj})
(\dot\gamma^i_\alpha)^k
(\dot\gamma^i_\alpha)^j\ae 0\ \forall\, \gamma^i\in\gamma$ and
$\alpha$. Thus, by property (2), $\Gamma(q)^\ell_{kj}(x)\equiv
\Gamma(q')^\ell_{kj}(x)$ on a dense subset of $\Sigma$ as
$\overline{\bigcup\{\, \gamma^i(I)\mid \gamma^i
\in\gamma\,\}}\equiv\Sigma$ by (2). Hence, invoking
the continuity of $\Gamma(h)$ for $h=q,q'$, it follows at once that
$\Gamma(q)\equiv\Gamma(q')$ on $\Sigma$.
Now, with respect to local coordinate basis,
$\Gamma(q)^i_{kj}= \frac 12 q^{ih}( \partial_k q_{hj}+ \partial_j
q_{hk}-\partial_h q_{kj})$ (and likewise for $q'$); consequently, $q$
and $q'$ are related homothetically; that is, $\exists\, c>0$
constant such that $q'=cq$.\footnote{ Note
trivially that as $q,q'$ are
positive-definite, $c<0$ is not an admissible solution.} More generally,
$q,q'$ are related by some coordinate transformation, as is shown below.

Let $f:\Sigma\to\Sigma$ be a smooth diffeomorphism, where
$\Sigma=(\Sigma,q)$ and set $\Sigma_f=
f(\Sigma)\defn(\Sigma, (f^{-1})^*q)$. Clearly,
if $\gamma:I\to\Sigma$ is a $q$-geodesic, then $\gamma:I\to\Sigma_f$ is
an $(f^{-1})^*q$-geodesic in $\Sigma_f$ and conversely, by symmetry (as
isometries map geodesics into geodesics).
Hence, in view of these two
observations, each $\gamma\in\Cal M_\infty[\Gamma^+_2]$ is assigned to a
unique 3-geometry of $\Sigma$, where the space of {\it 3-geometries} is
defined to be the quotient space $\Cal Q=\Gamma^+_2/\Diff^+(\Sigma)$.
Recall that each element $[q]\in \Cal Q$ is defined by
$[q]=\{\, f^*q\mid f\in\Diff^+(\Sigma)\,\}$. Let $\pi_+:\Gamma^+_2\to
\Cal Q$ denote the natural projection. Then, $\pi_+$ is open
\cite{2, p. 317, \S 3.1} and $\Cal
Q$ is a second countable, metrizable space \cite{2, p. 318, theorem 1}.

As a converse remark, notice
that if $\Sigma$ were not separable or that $\gamma_q=\{\,
\gamma^i_q\mid i\in\Bbb N\,\}$ were not chosen to satisfy (2),
$\gamma_q$ need not uniquely determine $[q]\in\Cal Q$.
For want
of a better term, call $\Cal M_\infty[\Gamma^+_2]$ the space of {\it
piecewise geodesic $\aleph_0$-loops}.
Now, a suitable topology can be defined on this space. To do this,
let $L_\infty$ be the set of
affinely parametrized, piecewise geodesic loops in $\Sigma$ and let
$L_\Sigma^\infty$ denote the countably infinite set-theoretic product of
$L_\Sigma$. Define an equivalence relation $R_\Sigma\subset
L_\Sigma^\infty\times L_\Sigma^\infty$ by
$$R_\Sigma\defn \{\, (\gamma,\gamma')\subset L_\Sigma^\infty\times
L_\Sigma^\infty : [\gamma]=[\gamma']\,\},$$
where $[\eta]\defn\{\, \eta^i\mid i\in\Bbb N\,\}$ is just the set of
components of $\eta\defn(\eta^i)_{i=1}^\infty$. Let $\pi_\Sigma:
L_\Sigma^\infty\to M_\Sigma\defn L_\Sigma^\infty/R_\Sigma$ denote the
natural map. Then clearly, as a subset, $\Cal M_\infty[\Gamma^+_2]\subset
M_\Sigma$.

Now, let $M_\infty\subset L_\Sigma^\infty$ be a subset satisfying
\roster
\item"(i)" for each $\gamma\defn(\gamma^i)_{i=1}^\infty,\, \gamma^i\neq
\gamma^j\;\, \forall\, i\neq j$,
\item"(ii)" $\pi_\Sigma(M_\infty)=\Cal M_\infty[\Gamma^+_2]$.
\endroster
It is clear from the definition of $M_\infty$ that there exists a family
of subsets $M_\sigma\subset M_\infty$ such that
\roster
\item"(a)" $M_\infty=\bigcup_\sigma M_\sigma$,
\item"(b)" $M_\sigma\cap M_{\sigma'}=\varnothing\;\, \forall\,
\sigma\neq\sigma'$,
\item"(c)" $\pi_\Sigma|M_\sigma: M_\sigma\to \Cal M_\infty[\Gamma^+_2]$
is a (set-theoretic) bijection.
\endroster

Let $h_\sigma\defn \pi_\Sigma| M_\sigma$ and for each $\gamma\in \Cal
M_\infty[\Gamma^+_2]$, set $\gamma_\sigma= h_\sigma^{-1}(\gamma)\in
M_\sigma$.\footnote{The subscript $\sigma$ on $\gamma_\sigma$ will be
omitted should no confusion arise from the context.} The subsets
$M_\sigma$ admit suitable metrics to be constructed below. Firstly, fix
a finite atlas $\frak A$ on $\Sigma$ and let $\tilde d_\Omega$ be a
metric on $\Cal L_\Sigma$ compatible with its quotient topology.
Then, for any pair $\gamma,\eta
\in M_\sigma$, let $d_\sigma(\gamma,\eta)\defn \sup_i \tilde d_\Omega
(\gamma^i,\eta^i)+\sup_i \tilde d'_\Omega(\gamma^i,\eta^i)$, where
$$\tilde d'_\Omega(\gamma^i,\eta^i)\defn \ess\{\, \| D^k\gamma^i(t)-
D^k\eta^i(t)\| : t\in I,\, k\geqq 1\,\}$$
with sup running over all relevant (finite) charts $(U,\varphi)\in \frak
A$, ess denoting that the expression $\|D^k\gamma^i(t)- D^k\eta^i(t)\|$
is {\sl not} defined only on a {\sl finite} (possibly zero) set of
points in $I$ wherein $\gamma^i$ and $\eta^i$ are not differentiable,
and $D^k\gamma^i(t)$ denotes the $k$th differential of $\gamma^i$ at $t$
in abused notation. It is routine to verify that $d_\sigma$ is indeed a
metric. In all that follows, $M_\sigma$ will be endowed with the
$d_\sigma$-topology.
\medpagebreak

\remark{2.6. Remark} Let $\overline{\frak A}$ denote the maximal atlas
of $\Sigma$ and define a topology on $M_\sigma$ to be generated by
subbasic open sets $N_\varepsilon(\gamma; (U_{\alpha(i)},
\varphi_{\alpha(i)})_{i=1}^\infty, K)$ in $M_\sigma$ to be constructed
below, where $K\subset I$ is compact, $\gamma^i(K)\subset U_{\alpha(i)}$
and $(U_{\alpha(i)},\varphi_{\alpha(i)})\in \overline{\frak A}
\;\,\forall\, i$. Denote $\{\, \alpha(i)\mid 1\leqq i\leqq\infty\,\}$ by
$\alpha$ and $(U_{\alpha(i)},\varphi_{\alpha(i)})_i$ by
$(U,\varphi)_\alpha$ for notational simplicity, and let
$$\tilde d'_{\sigma\alpha K}(\gamma^i,\eta^i)\defn \{\, \|D^k
\varphi_{\alpha(i)}\circ\gamma^i(t)- D^k\varphi_{\alpha(i)}
\circ\eta^i(t)\| : t\in K,\, k\geqq 1\,\}$$
whenever $\gamma^i(K),\eta^i(K)\subset U_{\alpha(i)}\;\, \forall\, i$.
Then, for a fixed $\gamma\in M_\sigma$ such that $\gamma^i(K)\subset
U_{\alpha(i)}\;\,\forall\, i$, let $N_\varepsilon(\gamma;
(U,\varphi)_\alpha, K)\defn\{\, \eta\in M_\sigma \mid \tilde
d_{\sigma\alpha K}(\gamma,\eta) <\varepsilon,\, \eta^i(K)\subset
U_{\alpha(i)}\;\,\forall\, i\,\}$, where
$$\tilde d_{\sigma\alpha K}(\gamma,\eta)\defn \sup_i\tilde
d_\Omega(\gamma^i,\eta^i)+ \sup_i \tilde d'_{\sigma\alpha K}(\gamma^i,
\eta^i).$$
It can be shown that this topology is equivalent to the $\tilde
d_\sigma$-topology on $M_\sigma$. In particular, the $\tilde
d_\sigma$-topologies on $M_\sigma$ defined relative to any two finite
atlases of $\Sigma$ are equivalent. Hence, in this sense, the
$\tilde d_\sigma$-topology is well-defined as it does not depend on the
choice of {\sl finite} atlas $\frak A$ on $\Sigma$.
\endremark
\medpagebreak

A topology on $\Cal M_\infty[\Gamma^+_2]$ can now be constructed.
Firstly, notice that the spaces $M_\sigma$ and $M_{\sigma'}$ are
homeomorphic for each pair $\sigma,\sigma'$---define $h_{\sigma\sigma'}:
M_\sigma\to M_{\sigma'}$ by $\gamma_\sigma\mapsto \gamma_{\sigma'}$,
where $h_\sigma(\gamma_\sigma)=\gamma= h_{\sigma'}(\gamma_{\sigma'})$.
The existence of $h_{\sigma\sigma'}$ follows immediately from conditions
(i) and (c). Hence, it is possible to endow $\Cal M_\infty[\Gamma^+_2]$
with a topology such that each $h_\sigma:M_\sigma\to \Cal
M_\infty[\Gamma^+_2]$ defines a homeomorphism. This will be the topology
imposed on $\Cal M_\infty[\Gamma^+_2]$. As an aside, if $M_\infty$ is
given the sum topology, $M_\infty\defn\bigoplus_\sigma M_\sigma$, then
$h: M_\infty\to \Cal M_\infty[\Gamma^+_2]$ given by $h|M_\sigma\defn
h_\sigma$ defines a continuous open surjection.
\bigpagebreak

\subheading{3. The Space of $\aleph_0$-Knots of $\Cal M_\infty[\Gamma_2^+]$}
\medpagebreak

First of all, some notations and
elementary properties
of the space of equivalence $\aleph_0$-loop
classes will be established. Let $\Cal
G_{\text{a}}^+$ be the set of (smooth) orientation-preserving,
ambient isotopies on $\Sigma$. That is,
$\Cal G_{\text{a}}^+\subset C^\infty(\Sigma\times I,
\Sigma\times I)$ is the following set:
$$\{\,F:\Sigma\times I\to\Sigma\times I\mid F(x,t)\defn (F_t(x),t),\,
F_0=\id_\Sigma,\, F_t\in\Diff^+(\Sigma)\;\,\forall\, t\in I\,\}$$
and define composition $\circ$ on $\Cal G^+_{\text{a}}$ by
$$(F'\circ F):(x,t)\mapsto (F'_t\circ F_t(x),t).$$
Then, clearly, $F'\circ F\in\Cal
G^+_{\text{a}}$ and $1_{\Sigma\times I}\defn \id_\Sigma\times\id_I\in\Cal
G^+_{\text{a}}$.
It is straight forward to check that $\langle\Cal
G^+_{\text{a}},\circ\rangle$ forms a group under $\circ$, where the
inverse $F^{-1}$ of $F=(F_t,\id_I)$ is defined to be $(F^{-1}_t,\id_I)$.
In particular, $\circ$ is compatible with the compact
$C^\infty$-topology on $\Cal G^+_{\text{a}}$---{\sl cf.} \cite{3, p. 64,
ex. 9}. Moreover, since $\Diff^+(\Sigma)$ is closed in
the group $\Diff(\Sigma)$ of smooth diffeomorphisms endowed with the
compact C${}^\infty$-topology (as it is a subgroup of
$\Diff(\Sigma)$),
$\Cal G^+_{\text{a}}$ is also
closed in $C^\infty(\Sigma\times I, \Sigma\times
I)$ (with respect to the compact C${}^\infty$-topology).

If $\gamma,\eta\in\Cal L_\Sigma$ are any pair of loops and $\gamma$ is
ambiently isotopic to $\eta$ under some $F\in\Cal G_{\text{a}}^+$,
denote this by $F:\gamma\simeq\eta$. Now, given any pair of
$\aleph_0$-loops $\gamma,\eta\in \Cal M_\infty[\Gamma_2^+]$,
define an equivalence relation $R$
generated by $\simeq$ on $\Cal M_\infty[\Gamma_2^+]$ as
follows:
$$\gamma\simeq\eta\quad\iff\quad\exists\, F\in\Cal G_{\text{a}}^+ \text{
such that } F\cdot\gamma=\eta,$$
where $F\cdot\gamma\defn\{\, F_1\circ\gamma^1,
F_1\circ\gamma^2,\dots\,\}$ and
$F:\gamma^i\simeq\eta^i\ \forall\,i$.
Then, the space $\Cal K[\Gamma_2^+]$ of equivalence classes
of $\aleph_0$-loops in $\Cal M_\infty[\Gamma_2^+]$ is defined to be
the quotient space $\Cal M_\infty[\Gamma_2^+]/\Cal G_{\text{a}}^+$.
Henceforth, for simplicity, the
term {\it (piecewise geodesic)
$\aleph_0$-knot} will mean an element of the quotient space
$\Cal K[\Gamma_2^+]$;
that is, an $\aleph_0$-knot denotes
an equivalence class of $\aleph_0$-loops under
a smooth, orientation-preserving, ambient isotopy.
The space $\Cal K[\Gamma_2^+]$ will be called the
{\it $(\aleph_0,\Gamma_2^+)$-knot space} of
$\Cal M_\infty[\Gamma_2^+]$. Let $\kappa_\infty:\Cal
M_\infty[\Gamma_2^+]\to \Cal K[\Gamma_2^+]$ denote the natural map,
where $\Cal K[\Gamma_2^+]$ is endowed with the quotient topology.

\proclaim{3.1. Lemma} The
natural projection $\kappa_\infty:\Cal M_\infty[\Gamma^+_2]\to\Cal
K[\Gamma^+_2]$ is
open.
\endproclaim
\demo{Proof} A sketch of the proof will be given.
To see that $\kappa_\infty$ is an open mapping, it is enough to
note that for each open subset $N\subset\Cal M_\infty[\Gamma_2^+]$,
$$\kappa_\infty^{-1}\circ\kappa_\infty(N)= \bigcup_{F\in\Cal
G^+_{\text{a}}} F\cdot N,$$
where $F\cdot N=\{\, F\cdot \gamma\mid \gamma\in N\,\}$.
Since
$F\cdot N$ is open in $\Cal M_\infty[\Gamma_2^+]$, as $F$ defines a
homeomorphism from $\Cal M_\infty[\Gamma_2^+]$ onto itself,
the quotient
topology implies that $\kappa_\infty^{-1}\circ \kappa_\infty(N)$, and
hence $\kappa_\infty$, must also be open. $\qed$
\enddemo

\proclaim{3.2. Proposition} $\Cal K[\Gamma_2^+]$ is Hausdorff.
\endproclaim
\demo{Proof} By lemma 3.1,
it will suffice to show that the equivalence relation
$R$ generated by
$\simeq$ is closed in $\Cal M_\infty[\Gamma^+_2]\times\Cal
M_\infty[\Gamma^+_2]$ \cite{4, p. 98, theorem 11}.
Let $\{(\gamma_n,\eta_n)\}_n$ be a sequence in
$R$ which converges in $\Cal
M_\infty[\Gamma_2^+]\times \Cal M_\infty[\Gamma_2^+]$
to $(\gamma_0,\eta_0)$. By
definition, $\exists$ a sequence
$\{F_n\}_n$ in $\Cal G_{\text{a}}^+$
such that $F_n:\gamma_n\simeq\eta_n$ for each $n$.
So, $(\gamma_n,F_n\cdot\gamma_n)\to (\gamma_0,\eta_0)\Rightarrow
F_n\cdot\gamma_n\to\eta_0$ and $\gamma_n\to\gamma_0$, and hence implying
that $\{F_n\}_n$ is a convergent sequence in $\Cal G^+_{\text{a}}$.
Consequently, $\Cal G^+_{\text{a}}$ is closed implies that $F_n\to
F_0\in\Cal G^+_{\text{a}}$ for some $F_0$. Whence, $\eta_0\equiv F_0\cdot
\gamma_0$ and $R$ is thus closed, as desired. $\qed$
\enddemo

In the interest of simplicity, call $\gamma\in\Cal M_\infty[\Gamma_2^+]$
a {\it piecewise $(\aleph_0,q)$-geodesic loop} whenever the 3-metric $q$
is required to be specified.

\proclaim{3.3. Lemma} Let $\gamma,\tilde\gamma\in
\Cal M_\infty[\Gamma_2^+]$ be
piecewise $(\aleph_0,q)$- and $(\aleph_0,\tilde q)$-geodesic
loops respectively. If
$\gamma\simeq\tilde\gamma$, then $\exists\, f\in\Diff^+(\Sigma)$ such
that $q=f^*\tilde q$.
\endproclaim
\demo{Proof} Let $F\in\Cal G^+_{\text{a}}$ be an ambient isotopy of
$\gamma$ and $\tilde\gamma$: $F\cdot\gamma=\tilde\gamma$.
Then, evidently, $\tilde\gamma$ is a piecewise
$(\aleph_0,(F_1^{-1})^*q)$-geodesic.
However, $\tilde\gamma$ is also a piecewise $(\aleph_0,\tilde
q)$-geodesic; hence, by \S 2 (2), $\exists\, f\in\Diff^+(\Sigma)$ such that
$\tilde q=f^*q$, as required. $\qed$
\enddemo
\bigpagebreak

\subheading{4. $\bold\aleph_0$-Knots and Classical Geometry}
\medpagebreak

In this section, the relationship between the equivalence classes of
$\aleph_0$-loops in $\Sigma$ and the
(classical) geometries admissible on $\Sigma$
will be studied. This correspondence can be easily sought simply by
noting that each element in $\Cal M_\infty[\Gamma_2^+]$ corresponds to
a unique 3-geometry $[q]$ of $\Sigma$ by construction.
The modified form
of Rovelli's Conjecture can now be formulated.

\proclaim{4.1. Theorem} There exists a
continuous, open surjection
$\hat\chi: \Cal
M_\infty[\Gamma^+_2]\to\Cal Q$ given by $\gamma_q\mapsto [q]$, where
$\gamma_q$ is a (piecewise) $(\aleph_0,q)$-geodesic loop and $q\in[q]$.
\endproclaim
\demo{Proof (Sketch)} Firstly, $\hat\chi$ is well-defined from the
definition of $\Cal M_\infty[\Gamma^+_2]$. Secondly,
the surjective property of $\hat\chi$ is also clear.
Thirdly, in this proof, $\Gamma^+_2$ will be
identified with its image under the (topological) imbedding $j^\infty:
\Gamma^+_2\hookrightarrow
C(\Sigma, J^\infty[p_\Sigma])$.\footnote{The notations
used here---the
C${}^\infty$-jets and compact C${}^\infty$-topology---can be found in
the appendix.} So, $\Cal Q\equiv j^\infty\Gamma^+_2/\Diff^+(\Sigma)$
and $\pi_+:j^\infty\Gamma^+_2\to\Cal Q$.

Now, fix some $\gamma_0\in\Cal
M_\infty[\Gamma^+_2]$ and let $N(q_0)= \bigcap_{i=1}^n M(K_i,
(\pi^{n_i}_\Sigma)^{-1}(U^{n_i}))$ be a neighbourhood of $q_0$ in
$\Gamma^+_2$, where $n_i\in\Bbb N$,
$n<\infty$ and $q_0\in\hat\chi(\gamma_0)=[q_0]$
is a representative of the $q_0$-equivalence class.
Set $N([q_0])=\pi_+(N(q_0))$.
Then, $\tilde N([q_0])\defn \pi_+^{-1}(N([q_0]))=
\bigcup\{\, f^*\circ
N(q_0)\mid f\in\Diff^+(\Sigma)\,\}$, where $f^*\circ N(q_0)\defn \{\,
f^*q\mid q\in N(q_0)\,\}$.
Let $D_\varepsilon(\gamma_0)$ be an $\varepsilon$-neighbourhood of
$\gamma_0$ defined by $B_\varepsilon(h^{-1}_\sigma(\gamma_0))=
h_\sigma^{-1}(D_\varepsilon(\gamma_0))\ \forall\, \sigma$.
Then, $\forall\, \eta\in D_\varepsilon(\gamma_0),\,
\tilde d_\Omega(
\gamma_{0\sigma}^i, \eta^i_\sigma)+
\tilde d'_\Omega(\gamma_{0\sigma}^i, \eta^i_\sigma)<
\varepsilon\ \forall\, i$ and $\sigma$, where
$h_\sigma^{-1}(\gamma)\defn \gamma_\sigma$.

Next, observe from the
definition that
$$(\ddot\gamma^i)^\ell + \Gamma (q)^\ell_{kj}(\dot
\gamma^i)^k (\dot\gamma^i)^j\ae 0\;\, \forall\, i\in\Bbb N\text{ and }
\ell=1,2,3,\tag$*$ $$
where $\Gamma(q)$ is a Riemannian connection determined by
the 3-metric $q$ (with the connection coefficients written with respect
to the natural frame for simplicity).
So, by choosing $\varepsilon>0$ to be sufficiently small,
and by fixing {\sl any} $\sigma$---and setting $\gamma_0^i=
\gamma^{\sigma(i)}_0$, $\eta^i=\eta^{\sigma(i)}$---it follows that
$|\ddot\eta^i-\ddot\gamma^i_0|< \varepsilon$ and
$|\dot\eta^i-\dot\gamma^i_0|<\varepsilon$
(almost everywhere), and in particular,
using $(*)$,
$$\split
{}&\phantom{\ae =}\!\Big| (\ddot\gamma^i_0)^\ell+ \Gamma(q_\eta)^\ell_{kj}
(\dot\gamma^i_0)^k (\dot\gamma^i_0)^j\Big|\\
&\ae\Big| (\ddot\eta^i+ \Cal O(\varepsilon))^\ell +
\Gamma(q_\eta)^\ell_{kj} (\dot\eta^i+ \Cal O(\varepsilon))^k (\dot\eta^i
+\Cal O(\varepsilon))^j\Big|\\
&\ae \Big|  (\ddot\eta^i)^\ell + \Gamma(q_\eta)^\ell_{kj}
(\dot\eta^i)^k (\dot\eta^i)^j +\Cal O(\varepsilon)\Big|\\
&\,\sim \,\Cal O(\varepsilon) \text{ a.e. on } I,
\endsplit$$
where $q_\eta\in\hat\chi(\eta)$.  Whence, $|(\ddot\gamma^i_0)^\ell
+\Gamma(q_\eta)^\ell_{kj} (\dot\gamma^i_0)^k (\dot\gamma^i_0)^j|\ae |
(\ddot\gamma^i_0)^\ell +\Gamma(q_0)^\ell_{kj}(\dot\gamma^i_0)^k
(\dot\gamma^i_0)^j- (\ddot\gamma^i_0)^\ell
-\Gamma(q_\eta)^\ell_{kj} (\dot\gamma^i_0)^k (\dot\gamma^i_0)^j|= |
(\Gamma(q_0)-\Gamma(q_\eta))^\ell_{kj}
(\dot\gamma^i_0)^k (\dot\gamma^i_0)^j|\sim
\Cal O(\varepsilon)$ a.e. (from above) $\forall\, i\in\Bbb N
\Rightarrow |\Gamma(q_\eta)^\ell_{kj}-
\Gamma(q_0)^\ell_{kj}|$ is small on $\Sigma$
for each fixed $\ell,\, k,\, j$
whenever $\varepsilon>0$ is small enough by appealing to \S 2 (2)
and the continuity of $\Gamma$. Thus,
from $\Gamma(q)^\ell_{kj}\defn \frac 12 q^{\ell h}( \partial_k q_{hj}+
\partial_j q_{hk}-\partial_h q_{kj})$ (in the natural frame), it follows
that $\exists\, f\in\Diff^+(\Sigma)$ such that
$f^*q_\eta$ and $q_0$, together with their $k$th derivatives,
must be close to one another: $f^*q_\eta(K_i)\subset
(\pi^{n_i}_\Sigma)^{-1}(U^{n_i})\ \forall\, i=1,\dots, n$.
So, $f^*q_\eta$ and hence $q_\eta$ must both belong
to $\tilde N([q_0])$ for $\varepsilon>0$ sufficiently small. Whence,
$\hat\chi(D_\varepsilon(\gamma_0))\subset N([q_0])$, and the
continuity of $\hat\chi$ follows.

Finally, to conclude this proof, observe that for any $\gamma\in\Cal
M_\infty[\Gamma^+_2]$, $\hat\chi^{-1}\circ\hat\chi(\gamma)= \{\,
f\circ\gamma\mid f\in\Diff^+(\Sigma)\,\}$, where $f\circ \gamma\defn\{\,
f\circ\gamma^1,f\circ\gamma^2,\dots\}$. Hence, for any
$\varepsilon$-neighbourhood $D_\varepsilon(\gamma)$,
$$\hat\chi^{-1}\circ \hat\chi(D_\varepsilon(\gamma))=
\bigcup_{f\in\Diff^+(\Sigma)}f\circ D_\varepsilon(\gamma),$$
and $\hat\chi$ is thus open, as desired. $\qed$
\enddemo

In spite of the
divergent approach given here with Rovelli's original idea, the
following corollary could perhaps be
christened as the {\it weak Rovelli conjecture}
inasmuch as the notion of relating knots with geometry originated
from Rovelli \cite{6}.

\proclaim{4.2. Corollary (Weak Rovelli Conjecture)}
The map $\hat\chi$ induces a
continuous, open surjection $\chi: \Cal
K[\Gamma_2^+]\to\Cal Q$ given by $[\gamma_q]\mapsto \hat\chi(\gamma_q)$,
where $\gamma_q\in\kappa_\infty^{-1}([\gamma_q])$ is any
fixed representative.
\endproclaim
\demo{Proof} This map
$\chi$ is well-defined by lemma 3.3. The result now follows
immediately from theorem 4.1, lemma 3.1
and the commutativity of the following
diagram:
$$\CD
\Cal M_\infty[\Gamma^+_2] @>\hat\chi>> \Cal Q\\
@V{\kappa_\infty}VV @VV{\id}V\\
\Cal K[\Gamma^+_2]@>{\chi}>> \Cal Q.
\endCD$$
$\qed$
\enddemo

Two comments regarding theorem 4.1 and its corollary are now in order.
Firstly, it is certainly evident that if $\Sigma$ be
separable (which,
here, it is in any case!), then it is sufficient to
characterized its 3-geometries by the
$\aleph_0$-loops in $\Cal M_\infty[\Gamma^+_2]$ since, by construction,
$\overline{\{\,\gamma^i(0)\mid i\in\Bbb N\,\}}\equiv \Sigma$, whereas
$n$-loops, for $n<\infty$ (using this construction), are not sufficient
to determine the 3-geometry uniquely (as might well be expected): {\sl
cf.} \S 2.5 for a detailed account.

Secondly, it has been established elsewhere---{\sl cf.} for example,
\cite{7, p. 132, \S 5.1} using the diffeomorphism constraints of general
relativity (in the loop representation)---that functionals on $\Cal
L_\Sigma$ which describe gravitational states are constant on the $\Cal
G_{\text{a}}^+$-orbits of $\Cal L_\Sigma$:
$\psi[\gamma]=\psi[\gamma']\;\,\forall\, \gamma,\gamma'\in [\gamma]$,
where $\psi:\Cal L_\Sigma\to\Bbb C$ is a loop functional.  However,
surprisingly, this condition follows immediately from corollary 4.2.
This can be easily seen as
follows. Functionals on $\Gamma^+_2$ that describe gravitational states
are those which are invariant under $\Diff^+(\Sigma)$:
{\sl i.e.}, they are essentially functionals on $\Cal Q$. Let
$C(\Cal Q,\Bbb C)$ be the set of continuous functionals on $\Cal Q$
and let $C(\Cal K[\Gamma_2^+],\Bbb C)$ be the set of
functionals on $\Cal K[\Gamma_2^+]$. Then, $\forall\,\tilde\Psi\in
C(\Cal Q,\Bbb C),\, \tilde\Psi\circ\chi\in C(\Cal
K[\Gamma_2^+],\Bbb C)$; that is, $\chi^*(C(\Cal Q,\Bbb C)) \subset
C(\Cal K[\Gamma_2^+],\Bbb C)$, and the assertion thus follows.

This concludes the
classical description of
$\aleph_0$-knots and their relationship with 3-geometries.
\bigpagebreak

\subheading{5. Discussion}
\medpagebreak

In this final section, a possible physical interpretation---albeit a
highly speculative one!---regarding knots and gravity will be sketched.
As was pointed out before, the separability of $\Sigma$ guarantees that
$\hat\chi$ in theorem 4.1 remains well-defined. Furthermore, as
classically, gravity---or equivalently, the 4-metric---of space-time is
determined by the distribution of matter in the universe via
Einstein's field equations, gravity is a `global' concept.  In this
sense, if $n$-loops can describe gravity in any way, then, provided that
space-time be separable, loops that will best describe it are
$\aleph_0$-loops. Indeed,
a judicious choice of $\aleph_0$-loops---such as
those given in the preceding sections---enables one to recover the
underlying Riemannian
3-manifold $\Sigma$ simply because $\overline{\{\, \gamma^i_q(0)\mid
i\in\Bbb N\,\}}=\Sigma$, and $\hat\chi(\gamma_q)=[q]$.  In the light of
this observation, it is not unreasonable to conclude that
gravity is the result of the way 3-space (and
hence, space-time) is knotted, where $(\Sigma,q)$ is said to be
{\it $[\gamma]$-knotted} if $\chi([\gamma])=[q]$.
And since $\chi$ is not one-one, $\Sigma$ can be knotted in
two $\Cal G_{\text{a}}^+$-inequivalent ways and yet give rise to the
same gravitational configuration (determined by $\chi$).
In short, having determined $\Cal M_\infty[\Gamma^+_2]$ from $\Sigma$,
each element in $\Cal M_\infty[\Gamma^+_2]$ contains the necessarily
information to reconstruct $\Sigma$.

To conclude with a speculative note on the quantum aspect of a knot
$[\gamma]$, one might heuristically interprete a {\it knot state}
$|[\gamma]\rangle$ to correspond to the pair $[(\Sigma, q)]$, where
$[(\Sigma,q)]\defn \{\, (\Sigma, q)\mid q\in \chi([\gamma])\,\}$. In
particular, $|[\gamma]\rangle$ is associated with a particular
3-geometry $\chi([\gamma])$.
Thus, $|[\gamma]\rangle$ corresponds to the global degrees of freedom of
gravity: and since gravitons are associated with the local degrees of
freedom of gravity, it has {\sl no direct} relationship with a knot
state.
In the full quantum theory, it is
quite reasonable to expect that $|[\gamma]\rangle$ will not span a
Hilbert space due to the highly non-linear nature of gravity
and the violation of the asymptotic completeness condition. Hence,
a knot state most probably
cannot be interpreted in the usual quantum field theoretic sense in that
it lies in some Hilbert space, although it is tempting to conjecture
that the knot states lie in some $\aleph_0$-dimensional smooth
K\"ahler manifold.
\bigpagebreak

\subheading{Acknowledgment}  The first author
thanks S. Scott, L. Tassie and  P. Leviton
for some fruitful conversations.
\bigpagebreak

\heading Appendix\endheading
\bigpagebreak

\subheading{A. Compact C$^\infty$-Topology}
\medpagebreak

The definition of a {\it compact
C${}^\infty$-topology} will be reviewed \cite{5, pp. 32--33, \S\S
4.1--4.3}. Let $J^n[\Sigma]$ be the space of $C^n$-jets from $\Sigma$
into $\Sigma$
and denote an element in $J^n[\Sigma]$ by either
$j^n(f(x))$ or $[f,x]_n$ (which ever proves more convenient).
Fix an atlas $\frak A_\Sigma=\{(U_\alpha,
\psi_\alpha)\}_{\alpha\in\Lambda}$ on $\Sigma$ and set $\frak
A_\Sigma(U_\alpha)=\{\, U\subset U_\alpha\mid U\text{ open }\}$. Then,
$\frak B_\Sigma=\bigcup_\alpha\frak A_\Sigma(U_\alpha)$ forms a base for
$\Sigma$. Let $J^0[\Sigma]=\Sigma\times\Sigma$ and let $\pi^0_1:
J^1[\Sigma]\to J^0[\Sigma]$ by $j^1\phi(p)\mapsto (p,\phi(p))$. Set
$U_{\alpha\alpha'}^1\equiv (\pi^0_1)^{-1}(U_\alpha\times U_{\alpha'})$
and define $p_{\pm}^1:J^1[\Sigma]\to\Sigma$ by $p_+^1:j^1\phi(p)\mapsto
\phi(p)$ and $p^1_-: j^1\phi(p)\mapsto p$. Then, it is clear that
$U_{\alpha\alpha'}^1= (p^1_-)^{-1}(U_\alpha)\cap
(p^1_+)^{-1}(U_{\alpha'})$.  Finally, let $\frak A^1_{\alpha\alpha'}=
\{\, (\pi^0_1)^{-1}(U\times U')\mid U\times U'\subset U_\alpha\times
U_{\alpha'}\text{ open }\}$. Then, $\frak B^1= \bigcup_{\alpha, \alpha'}
\frak A_{\alpha\alpha'}^1$ forms a base for $J^1[\Sigma]$.
Following \cite{8, p. 94, definition 4.1.5}, define
$\Psi^1_{\alpha\alpha'}: U^1_{\alpha\alpha'}\cong
{}^3B_{\varepsilon_\alpha}(x_\alpha)\times
{}^3B_{\varepsilon_{\alpha'}}(x_{\alpha'})\times
{}^{N_1}B_{\varepsilon_1}(x_1)$ by
$$[\phi,p]_1\mapsto (\psi_\alpha(p),\psi_{\alpha'}(\phi(p)),
D_\alpha(j^1\phi(p))),$$
where ${}^nB_\varepsilon(x)$ is an open $\varepsilon$-ball in $\Bbb R^n$
and
$D_\alpha j^1\phi(p)\defn \lbr\frac{\partial}{\partial x^i_\alpha}
\phi_{\alpha\alpha'}(\psi_\alpha(p))\rbr_i$
for some $N_1\in\Bbb N$
such that $D_{\alpha\alpha'}: U^1_{\alpha\alpha'}\cong
{}^{N_1}B_{\varepsilon_1}(x_1)$ and $\phi_{\alpha\alpha'}\defn
\psi_{\alpha'}\circ \phi\circ\psi_\alpha^{-1}$.
The pair
$(U^1_{\alpha\alpha'}, \Psi_{\alpha\alpha'}^1)$ defines a chart on
$J^1[\Sigma]$. Denote $\Psi^1_{\alpha\alpha'}$
symbolically by $\psi_\alpha\times\psi_{\alpha'}\times D_\alpha$.

Now, define $p^2_{\pm}: J^2[\Sigma]\to\Sigma$ by $p^2_-:
j^2\phi(p)\mapsto p$ and $p^2_+: j^2\phi(p)\mapsto \phi(p)$.
Furthermore, define $\pi^0_2: J^2[\Sigma]\to J^0[\Sigma]$ by $\pi^0_2(
j^2\phi(p))=(p,\phi(p))$ and let $U^2_{\alpha\alpha'}\defn
(\pi^0_2)^{-1}(U_\alpha\times U_{\alpha'})$. Then,
$U^2_{\alpha\alpha'}\equiv (p^2_-)^{-1}(U_\alpha)\cap
(p^2_+)^{-1}(U_{\alpha'})$. And as
with the case for $J^1[\Sigma]$, the pair $(U^2_{\alpha\alpha'},
\Psi^2_{\alpha\alpha'})$ defines a chart in $J^2[\Sigma]$, where
$\Psi^2_{\alpha\alpha'}\defn \psi_{\alpha}\times \psi_{\alpha'}\times
D_\alpha\times D_\alpha^2$ and $D^2_\alpha: U^2_{\alpha\alpha'}\cong
{}^{N_2}B_{\varepsilon_2}(x_2)$, for some $\varepsilon_2>0$ and some
$N_2\in\Bbb N$, is defined by $D^2_\alpha(j^2\phi(p))\defn \lbr
\frac{\partial^2}{\partial
x^{i_1}_\alpha\partial x^{i_2}_\alpha}
\phi_{\alpha\alpha'}(\psi_\alpha(p))\rbr_{i_1\leqq i_2}$.
Also, define $\pi^1_2:J^2[\Sigma] \to
J^1[\Sigma]$ by $[\phi,p]_2\mapsto [\phi,p]_1$. Then, by definition,
$\pi^0_2=\pi^0_1\circ \pi^1_2$ and $\pi^1_2(U^2_{\alpha\alpha'})=
U^1_{\alpha\alpha'}$.\footnote{ For $j^1\phi(p)\in
U^1_{\alpha\alpha'}\Rightarrow \pi^0_1(j^1\phi(p))=(p, \phi(p))\in
U_\alpha\times U_{\alpha'}\Rightarrow j^2\phi(p)\in U^2_{\alpha\alpha'}$
and so, $ U^1_{\alpha\alpha'}\subseteq
\pi^1_2(U^2_{\alpha\alpha'})$. Conversely, $j^1\phi'(p')\in
\pi^1_2(U^2_{\alpha\alpha'})\Rightarrow j^2\phi'(p')\in
U^2_{\alpha\alpha'}\Rightarrow (p',\phi'(p'))\in U_\alpha\times
U_{\alpha'}\Rightarrow$ the converse set-inequality, as required.}
Finally, let $\frak A^2_{\alpha\alpha'}=\{\, (\pi^0_2)^{-1} (U\times
U')\mid U\times U'\subset U_\alpha\times U_{\alpha'} \text{ open }\}$;
then, $\frak B^2=\bigcup_{\alpha,\alpha'}\frak A^2_{\alpha\alpha'}$ forms
a base for $J^2[\Sigma]$.

By induction, given $J^n[\Sigma]$,
$(\pi^0_n)^{-1}(U_\alpha\times U_{\alpha'})= (p^n_-)^{-1}(U_\alpha)\cap
(p^n_+)^{-1}(U_{\alpha'})$ and $\pi^{n-1}_n(U^n_{\alpha\alpha'})=
U^{n-1}_{\alpha\alpha'}$.
Furthermore, the pair
$(U^n_{\alpha\alpha'},\Psi^n_{\alpha\alpha'})$ forms a chart on
$J^n[\Sigma]$ as follows: $\Psi^n_{\alpha\alpha'}\defn \psi_\alpha\times
\psi_{\alpha'}\times \prod_{i=1}^nD^i_\alpha$, where
$$D^\ell_\alpha: [\phi,p]_\ell\mapsto \lbr \frac{\partial^\ell
\phi_{\alpha\alpha'}\circ\psi_\alpha(p)}{\partial x_\alpha^{i_1} \dots
\partial x_{\alpha}^{i_\ell}} \rbr_{i_1\leqq\dots\leqq i_\ell}\in \Bbb
R^{N_\ell}$$
with some $N_\ell\in\Bbb N$ such that
$D^\ell_\alpha(U^\ell_{\alpha\alpha'})={}^{N_\ell}B_{\varepsilon_\ell}
(x_\ell)$. Tersely, $\Psi^n_{\alpha\alpha'}: U^n_{\alpha\alpha'}\cong
{}^3B_{\varepsilon_\alpha}(x_\alpha)\times {}^3B_{\varepsilon_{\alpha'}}
(x_{\alpha'})\times\prod_{i=1}^n {}^{N_i}B_{\varepsilon_i}(x_i)$. The
topology on $J^n[\Sigma]$ is generated by the base $\frak B^n=
\bigcup_{\alpha\alpha'}\frak A^n_{\alpha\alpha'}$, where $\frak
A^n_{\alpha\alpha'}=\{\, (\pi^0_n)^{-1}(U\times U')\mid U\times
U'\subset U_\alpha\times U_{\alpha'}\text{ open }\}$.

It follows from the construction that $\{J^n[\Sigma],\pi^{n-1}_n,\Bbb
N\}$ forms an inverse sequence. Let $J^\infty[\Sigma]\defn\varprojlim
J^n[\Sigma]$ denote the limit of the inverse sequence. Then, $\frak
B^\infty=\{\, (\pi^n)^{-1}(U)\mid U\in \frak B^n\ \forall\, n\,\}$
defines a base of $J^\infty[\Sigma]$, where $\pi^n\defn p^n|
J^\infty[\Sigma]$ and $p^n:\prod_{i\in\Bbb N}J^i[\Sigma]\to
J^n[\Sigma]$ is the $n$th projection. Observe from \cite{1, p. 98,
proposition 2.5.1} that $J^\infty[\Sigma]$ is closed in the Cartesian
product $\prod_{i\in\Bbb N} J^i[\Sigma]$.

The {\it compact} (or {\it weak}) {\it C${}^\infty$-topology} on
$C^\infty(\Sigma,\Sigma)$ is the topology induced by the map
$j^\infty: C^\infty(\Sigma,\Sigma)\to C(\Sigma, J^\infty[\Sigma])$
defined by $f\mapsto j^\infty f\defn [f,\cdot\,]_\infty$ such
that it is a topological imbedding.
Let $\Diff(\Sigma)\subset C^\infty(\Sigma,
\Sigma)$ denote the set of C${}^\infty$-diffeomorphisms on $\Sigma$.
The composition mapping $\circ :
\Diff(\Sigma)\times\Diff(\Sigma)\to \Diff(\Sigma)$ given by
$(f,g)\mapsto f\circ g$ defines a group structure on $\Diff(\Sigma)$.
Indeed, the group structure is compatible with the compact
C${}^\infty$-topology on $\Diff(\Sigma)$ \cite{3, p. 64, ex. 9}.
Lastly, observe from \cite{3, p. 38, theorem 1.6} that $\Diff(\Sigma)$
is open in $C^\infty(\Sigma,\Sigma)$ (as $\Sigma$ is compact implies
that the weak
and strong C${}^\infty$-topology coincide).

This appendix will conclude with a brief sketch of the compact
C${}^\infty$-topology on the space $\Gamma^+_2$ of
(admissible) Riemannian metrics on $\Sigma$. Let $p_\Sigma :
S^+_2\Sigma\to\Sigma$ be the symmetric covariant 2-tensor bundle over
$\Sigma$ and $p_{\Sigma n}: J^n[p_\Sigma]\to\Sigma$
be the C${}^n$-jet bundle of
the cross-sections of $S^+_2\Sigma$. Then, defining $\pi^0_{\Sigma 1}:
J^1[p_\Sigma]\to \Sigma\times S^+_2\Sigma$ as above by $j^1q(x)\mapsto
(x,q(x))$ and $\pi^m_{\Sigma n}: J^n[p_\Sigma]\to J^m[p_\Sigma]$ by
$j^nq(x)\mapsto j^mq(x)$ whenever $m\leqq n$, one again obtains an
inverse sequence $\{J^n[p_\Sigma], \pi^{n-1}_{\Sigma n},\Bbb N\}$,
where $J^0[p_\Sigma]\defn \Sigma\times S^+_2\Sigma$. Finally, let
$J^\infty[p_\Sigma]$ denote the inverse limit of the sequence and set
$\pi_\Sigma^n\defn p^n_\Sigma| J^\infty[p_\Sigma]$, where $p^n_\Sigma:
\prod_{i\in\Bbb N} J^i[p_\Sigma]\to J^n[p_\sigma]$ is the
$n$th projection. The
topology of $\Gamma^+_2$ is then defined by the (topological) imbedding
$j^\infty:\Gamma^2_+\hookrightarrow C(\Sigma, J^\infty[p_\Sigma])$.
\bigpagebreak


\Refs
\ref
\no 1
\by Engelking, R.
\book General Topology
\publ SSPM 6, Heldermann Verlag-Berlin
\yr 1989
\endref

\ref
\no 2
\by Fischer, A. E.
\paper Theory of Superspace
\jour Relativity, ed. Carmeli, Fickler, Witten
\endref


\ref
\no 3
\by Hirsch, M.
\book Differential Topology
\publ GTM 33, Springer-Verlag, Berlin
\yr 1976
\endref

\ref
\no 4
\by Kelley, J.
\book General Topology
\publ Van Nostrand, New York
\yr 1955
\endref

\ref
\no 5
\by Michor, P.
\book Manifolds of Differentiable Mappings
\publ Shiva Maths Series, Kent
\vol 8
\yr 1980
\endref

\ref
\no 6
\by Rovelli, C.
\jour Class. Quantum Grav
\vol 8
\yr 1991\pages 1613--1675
\endref

\ref
\no 7
\by Rovelli, C. and Smolin, L.
\jour Nuc. Phys
\vol B331\yr 1990\pages 80--152
\endref

\ref
\no 8
\by Saunders, D.
\book The Geometry of Jet Bundles
\publ LMS 142, Cambridge University Press
\yr 1989
\endref
\endRefs
\enddocument